\documentclass[pra,reprint,showpacs,twocolumn,superscriptaddress,floatfix]{revtex4-1}

\usepackage{epsfig}
\usepackage{amsmath}
\usepackage{amssymb}
\usepackage{amsfonts}
\usepackage{mathptmx}
\usepackage{dcolumn}
\usepackage{eucal}
\usepackage{bm}
\usepackage{color}
\usepackage[colorlinks,linkcolor=blue,citecolor=blue]{hyperref}
\usepackage{epstopdf}
\usepackage{bbold}
\usepackage{url}
\usepackage{mathtools}
\usepackage{dsfont}

\newcommand{\Trace}{{\rm Tr}}

\def \ket#1{\mathinner{|{#1}\rangle}}

\newcommand{\matrixel}[3]{{\mathinner{\langle{#1}| {#2} | {#3}\rangle}} }

\newcommand{\GFunc}{\mathcal{G}_\lambda}

\def \Ham{\hat{H}}

\def \QNDM{QNDM~}
\def \DM{DM~}

\def \LiH{LiH~} 
\def \Li{Li$_2$}
\def \H2{H$_2$}
\def \OH{OH~}

\begin{document}

\title{Resource reduction for variational quantum algorithms by non-demolition measurements}

\author{D. Melegari}
\email[]{These authors  contributed  equally to this work.}
\affiliation{Dipartimento di Fisica, Universit\`a di Genova, via Dodecaneso 33, I-16146, Genova, Italy}
\affiliation{INFN - Sezione di Genova, via Dodecaneso 33, I-16146, Genova, Italy}
\author{R. Abdul Razaq}
\email[]{These authors  contributed  equally to this work.}
\affiliation{Dipartimento di Fisica, Universit\`a di Genova, via Dodecaneso 33, I-16146, Genova, Italy}
\affiliation{INFN - Sezione di Genova, via Dodecaneso 33, I-16146, Genova, Italy}
\author{G. Minuto}
\affiliation{Dept. of Informatics, Bioengineering, Robotics, and Systems Engineering (DIBRIS), Polytechnic School of Genoa University, Genova, Italy}
\affiliation{INFN - Sezione di Genova, via Dodecaneso 33, I-16146, Genova, Italy}
\author{P. Solinas}
\affiliation{Dipartimento di Fisica, Universit\`a di Genova, via Dodecaneso 33, I-16146, Genova, Italy}
\affiliation{INFN - Sezione di Genova, via Dodecaneso 33, I-16146, Genova, Italy}

\def\thefootnote{*}\footnotetext{These authors contributed equally to this work}

\date{\today}

\begin{abstract}
We present a comparative study of two implementations of a variational quantum algorithm aimed at minimizing the energy of a complex quantum system.
In one implementation, we extract the information of the energy gradient by projective measurements.
In the second implementation, called the Non-Demolition approach, the gradient information is stored in a quantum detector, which is eventually measured.
As prototypical examples, we study the energy minimization of the Lithium-based molecules and then extend the analysis systems with increased complexity.
We find that, while both approaches are able to identify the energy minimum, the Non-Demolition approach has a clear advantage in terms of the overall computational resources needed.
This advantage increases linearly with the complexity of the quantum system, making the Non-Demolition approach the ideal candidate to implement such variational quantum algorithms.
\end{abstract}

\maketitle


\section{Introduction}
\label{sec:Introduction}

Quantum systems are notoriously hard to simulate in classical computers \cite{feynman1982}.
This is particularly true of many-body quantum systems that are composed of a huge number of atoms and electrons.
In this case, the quantum correlations and entanglement cannot be neglected and it is impossible reduce the simulation space which increases exponentially with the number of particles.
As a consequence, the classical computational memory is quickly saturated. This poses a strong limitation on the dimension of quantum systems, which can be simulated with classical hardware. 
Indeed, the simulations of biological molecules and high-temperature superconductors remain out of the reach of even the most powerful classical computers.

In this framework, quantum computers might help us reduce the time and cost of these simulations and make them accessible in the future.
Because they exploit the laws of quantum mechanics, the problems of simulation dimensions and correlation complexity are immediately solved.

Despite the fact that it is not yet clear for which problems we can have a quantum advantage, there are a few categories which are believed to be well-fitted for the quantum simulations.
Among these, the simplest is to determine the minimal energy and the ground state of a quantum system.
Since physical systems naturally evolve towards their ground state, in the absence of perturbations or in the low energy/temperature case, it is one in which the system is mostly found.
Then, knowing the properties of the ground state gives us invaluable information about its behaviour and properties.

Starting from these ideas, the most straightforward algorithms to address this problem are borrowed from the variational algorithms, and are named Variational Quantum Algorithm (VQA) (see Ref. \cite{Peruzzo_2014} and Sec. \ref{sec:VQA}).
The commonly used approaches to VQA are hybrid quantum algorithms.
The complex calculation of the system properties is performed with the quantum computers; the data are then passed to a classical computer which performs a simpler and straightforward step.
More specifically, the quantum computer is used to extract  information about the gradient of a cost function (energy) in a point of the parameter space of the system.
Given this, the classical algorithm decides how to change the system parameters in order to minimize the cost function.
Iterating this procedure leads to a (local or global) minimum of the cost function.   

Traditionally, the measurement of the gradient has been done through projective or Direct Measurements (\DM) \cite{schuld2015, McClean_2016, McClean2018,SchuldPRA2019, Mari2021, Cerezo2021}.
However, this approach is time and resource-consuming since it requires running the quantum computational circuits many times.
To reduce the quantum computational resources, recently, an alternative way to extract the gradient information has been proposed 
\cite{Solinas_2023, minuto2024novel}.
This is based on the Quantum non-demolition measurement (\QNDM) \cite{Braginsky1980, BraginskyBook, Caves1980,  solinas2015fulldistribution,solinas2016probing,solinas2021,solinas2022} and exploits a quantum detector to store the information about the average derivative of the energy at a point.
This considerably reduces the resources needed to measure the gradient and run the VQA \cite{minuto2024novel,Solinas_2023}.
At the moment, this approach has been analyzed only in terms of gradient measurement, but its applicability to the optimization procedure has even been discussed.

In this article, we fill this gap by comparing the efficiency of the  \DM and \QNDM  approaches to the optimization procedure.
As a test-bench, we use the problem of finding the minimum energy of some real (although approximated) molecules: \H2 and \LiH.
We show that, while both algorithms converge to the energy minimum within a reasonable time, the \QNDM  approach needs fewer resources, making it the best choice for the minimization task.

Then, we extend the study to minimize the energy of another (but still physical) system. This allows us to increase the Hamiltonian complexity keeping a sound statistical analysis.
This allows us to compare the two approaches in a more realistic situations.
In these case, the \QNDM  advantage in terms of quantum computational logical gates increases further with respect to the  \DM approach.

The article structure is the following.
In Sec. \ref{sec:VQA}, we present the key ideas of the VQA, 
and in Sec. \ref{sec:DM_vs_QNDM}, we discuss the two approaches to measure the gradient of the cost function.
Section \ref{sec:simulations} contains the details of the numerical simulations and the results.
The conclusions are drawn in Sec. \ref{sec:conclusions}.


\section{Variational quantum algorithms}
\label{sec:VQA}

In this section, we recap the fundamentals of Variational Quantum Algorithms (VQAs) \cite{Peruzzo_2014,Solinas_2023,Cerezo2021}. The problem we want to solve is the following: given a complex Hamiltonian $\Ham$, find the minimum energy. 

In many practical scenarios, the Hamiltonian $\Ham$ in a quantum computer implementation is composed of a sum of Pauli strings $\hat{P}_i = \prod_j \hat{A}_i^j$, where $\hat{A}_j \in [X_j, Y_j, Z_j, I_j]$, and $X_j, Y_j, Z_j$ are the usual Pauli operator acting on the $j$-th qubit \cite{McArdle2020,TILLY20221}. In this case, the Hamiltonian can be written as:
\begin{equation}
    \hat{H} = \sum_{i=1}^J h_i \hat{P}_i
    \label{eq:M}
\end{equation}
where $J$ is the total number of Pauli string composing $\hat{H}$, and $h_i$ is the weight associated to the $i$-th string.
The increase in the working space dimension, i.e., the number of qubits used, implies a more accurate model.

For an $n$ qubits system, initialized in the state $|\psi_0 \rangle = |0\rangle_1\otimes|0\rangle_2 \otimes\ldots \otimes |0\rangle_n \equiv |00...0 \rangle$, the generic unitary transformation acting on the $n-$ qubit system is denoted as $U({\vec \theta})$, where $\vec \theta$ is a vector in the parameter space, and the quantum state $|\psi({\vec \theta})\rangle$ is obtained by applying the transformation $U({\vec \theta})$ to the initial state $|\psi_0\rangle$, i.e. $|\psi(\vec{\theta})\rangle = U(\vec{\theta}) |\psi_0 \rangle$.

The unitary operator $U({\vec \theta})$ can be implemented as a sequence of $L$ parameterized unitary transformations $U_j({\theta_j})$, interleaved with unparametrized entangling transformations $V_j$, with $j\in[1, L]$, such that:
\begin{equation}
 	U({\vec \theta}) = V_L U_L({ \theta_L}) \dots V_2 U_2({ \theta_2}) V_1 U_1({\theta_1})\,.
	\label{eq:U_def}
\end{equation}
For this discussion, we'll consider $U_j({\theta_j}) = \exp{\{- i \theta_j \hat{H}_j \}}$, where $\hat{H}_j$ is the corresponding generator of the transformation. If $\hat{H}_j$ is composed by the Pauli operators, $\hat{H}_j^2 = \mathds{1}$, and we have that 
\begin{equation}
	U_j( \theta_j) = e^{- i \hat{H}_j \theta_j/2} = \cos \frac{\theta_j}{2} \mathds{1} -i \sin \frac{\theta_j}{2} \hat{H}_j.
    \label{eq:rot_par}
\end{equation}
This accounts for many relevant cases, such as when $H_j$ is a tensor product of any multi-qubit Pauli matrices \cite{Mari2021}. 

\begin{figure}
    \centering
    \includegraphics[width=1\linewidth]{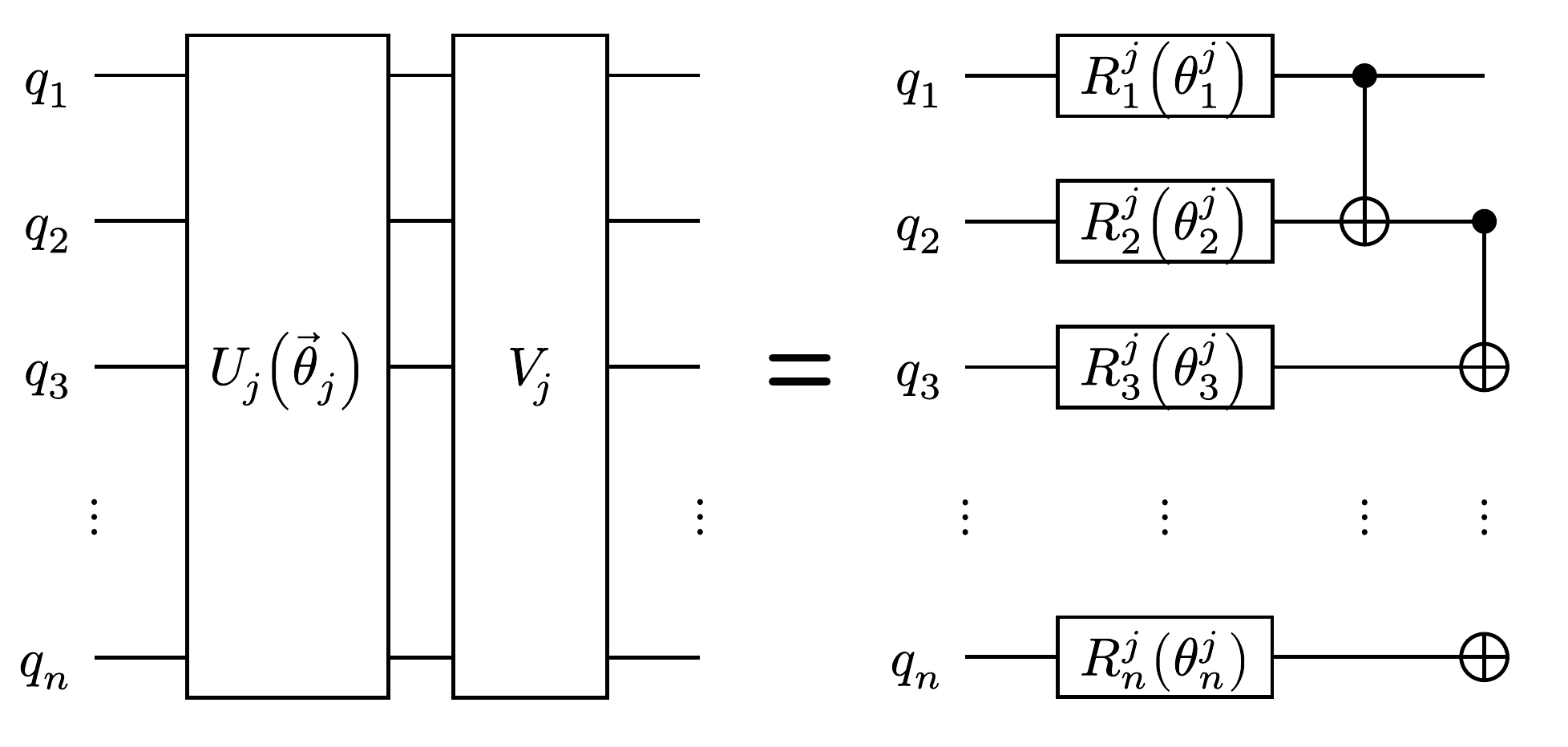}
    \caption{Schematic diagram of the $j-$th layer for an $n-$qubit circuit, as described in Eqs. \ref{eq:rot_layer} and \ref{eq:ent_layer}. In the left, we see the generic structure of a single layer of the circuit, while on the right we have the specific structure of the $j-$th layer adopted in this paper, for which the layer is composed of a sequence of parametrized single qubit gates $R^j_i(\theta^j_i)$ and entangling two-qubit gates $C_{i-1}NOT_i$.}    
    \label{fig:layer}
\end{figure}

However, to further simplify the discussion and the simulation, we restrict our attention to the case in which 
\begin{equation}\label{eq:rot_layer}
    U_j(\vec \theta_j) = R^j_1(\theta^j_1)\dots R^j_n(\theta^j_n)\,,
\end{equation}
where each operator $R^j_i(\theta^j_i)$ is a parameterized single qubit rotational gate (see Eq.  (\ref{eq:rot_par})) and $\hat{H}_j$ is a single Pauli matrix. Likewise, for the $V_j$ part of the transformation, which is independent of the parameters, we select the following specific structure
\begin{equation}\label{eq:ent_layer}
    V_j = C_1NOT_2\dots C_{n-1}NOT_n\,,
\end{equation}
where $C_{i-1}NOT_i$ is a two-qubit operator with the control on the $i-1$-th qubit and the action on the $i$-th
qubit \cite{nielsen-chuang_book}.
The sequence of $U_j(\vec \theta_j)$ and $V_j$ is usually called {\it layer} and, as shown in Fig. \ref{fig:layer}, is a combination of parametrized and entangling unitary transformations. The number of layers, $L$, determines the depth of the circuit and is a crucial parameter in the optimization process. In the following, it will be useful to define the total number of gates to implement $U(\vec{\theta})$ as $k$. For this architecture, the value of $k$ depend on the number of qubits of the system, and the number of layers, and follows the relation $k = (2n-1)L$.

In optimization problems, the goal is to find the quantum state $\ket{\psi(\vec\theta)}$ (i.e. the best set of parameters $\vec{\theta}$) which minimizes the expectation value of the Hamiltonian $\Ham$.
To accomplish this task, we define a cost function $f$ that we want to minimize with respect to the parameters ${\vec \theta}$
\begin{equation}
	f({\vec \theta}) = \matrixel{\psi ({\vec \theta})}{\Ham}{\psi ({\vec \theta})} = \matrixel{0}{ U^\dagger({\vec \theta}) \Ham U({\vec \theta})}{0}\,.
	\label{eq:f_def}
\end{equation}

We adopt the so-called hybrid quantum-classical optimization procedure described in the introduction \cite{McClean_2016, Cerezo2021}.
The minimization algorithm consists of two steps that must be iterated until the minimum of $f({\vec \theta})$ is reached.
First, with the help of a quantum computer, we evaluate the gradient of $f({\vec \theta})$ in a given point of the parameter space $\vec \theta$. Then, the information about the gradient is fed into a classical computer which, with standard optimization algorithms \cite{kingma2017adam,Kubler2020adaptiveoptimizer, Sweke2020stochasticgradient}, evaluates in which direction of the $\vec \theta$ space we move to reach the minimum of $f$.
In this paper, we focus on the quantum part of the algorithm, that is, the calculation of the gradient of $f({\vec \theta})$ done with a quantum computer, and we do not discuss the classical one.

To measure the derivative of $f({\vec \theta})$ along the $j$-th direction in the parameter space $\vec \theta$, we measure $f(\vec \theta +s \hat e_j )$ and $f(\vec \theta -s \hat e_j )$ where $\hat e_j$ is the unit vector along the $\theta_{j}$ direction and $s$ is a shift parameter.
Then, we calculate the quantity \cite{Mari2021}
\begin{equation}
	g_{j} = \frac{\partial f({\vec \theta})}{\partial \theta_{j}} = \frac{f(\vec \theta +s \hat e_j ) - f(\vec \theta -s \hat e_j)}{ 2 \sin s }\,.
	\label{eq:f_grad_def}
\end{equation}

With $s=\pi/2$, we obtain the parameter shift rule described in Refs. \cite{Li2017,Mitarai2018,SchuldPRA2019,Mitarai2019}.
Notice that $g_{j}$ is exactly the derivative of $f({\vec \theta})$ with no approximation and repeating the procedure for all the directions $\theta_j$ we obtain the gradient of $f({\vec \theta})$, which is then fed into the classical optimization algorithm.

For this purpose, we employed the standard gradient descent algorithm \cite{ruder2017overviewgradientdescentoptimization}, where the parameters are iteratively updated according to the following rule:
\begin{equation}\label{eq:gradient_descent}
    \vec{\theta}_{\text{new}} = \vec{\theta}_{\text{old}} - \eta~ \nabla f(\vec{\theta}_{\text{old}}),
\end{equation}
where $\eta$ represents the learning rate. The selection of an appropriate $\eta$ is critical for the algorithm's convergence properties. Specifically, an excessively large learning rate may lead to rapid convergence but risks overshooting the optimal solution, while a smaller one may ensure more stable convergence at the cost of slower progress.

In the context of quantum optimization, the choice of $\eta$ must be carefully calibrated with respect to the error in the gradient estimation process. This is because the precision of the gradient measurement, directly influences the stability and efficiency of the optimization procedure. Thus, the learning rate must be adapted to balance the trade-off between convergence speed and the noise introduced by finite sampling.


\section{Direct measurement versus quantum non-demolition}
\label{sec:DM_vs_QNDM}

\subsection{Direct measurement}
\label{sec:DM}

The standard approach to measure the value of the derivatives of the cost function (\ref{eq:f_grad_def}) is the \DM method \cite{McClean_2016, McClean2018,SchuldPRA2019, Mari2021, Cerezo2021}.
This  consists of measuring the average values of the operator $\Ham$ in the two points, i.e., $\vec \theta \pm s \hat e_j$, and then calculating the derivative as in Eq. (\ref{eq:f_grad_def}).

Rephasing this scheme in the language and with the definition of Sec. \ref{sec:VQA}, to calculate the values of $f(\vec \theta +s \hat e_j )$, iwe implement the corresponding $U(\vec \theta +s \hat e_j )$ transformation, and then we perform a projective measurement {\it for each Pauli string} $\hat{P}_i$ that appears in the definition (\ref{eq:M}) of the observable $\Ham$ \cite{McArdle2020, Mari2021}.
For every Pauli string $\hat{P}_i$ in $\Ham$, we have to iterate these steps until the desired statistical accuracy is reached.
The same procedure is then implemented for $f({\vec \theta -s \hat e_j })$, and then, the derivative can be calculated as in Eq. (\ref{eq:f_grad_def}).

Following this procedure, a single derivative $g_j$ in the  \DM approach can be written as \cite{minuto2024novel}
\begin{eqnarray}
    g_j &=& \sum_i \frac{h_i}{2 \sin s}\Big( \Trace_S \big[\hat{P}_iU^{\dagger}(\vec \theta+s \hat e_j)\rho_s^0U(\vec \theta+s \hat e_j) \big] \nonumber \\
    && -\Trace_S \big[\hat{P}_iU^{\dagger}(\vec \theta-s \hat e_j)\rho_s^0 U(\vec \theta-s \hat e_j) \big]\Big)\,,
    \label{eq:DM}
\end{eqnarray}
where $\rho_s^0 =|\psi_0\rangle\langle \psi_0|$, $U(\vec \theta)$ is the operator in Eq. (\ref{eq:U_def}) and $\Trace_S$ denotes the trace over all the qubits. 


\subsection{Quantum non-demolition approach}
\label{sec:QNDM}

The key idea of the \QNDM method is to store the information about the derivative of the cost function in the phase of an ancillary qubit named the {\it detector}.
More specifically, by coupling the original system and the detector twice, it is possible to store the information about both $f(\vec \theta +s \hat e_j )$ and $f(\vec \theta -s \hat e_j )$ in the phase of the detector.
Then, the detector is measured to extract information about the derivative \cite{Solinas_2023}.

The advantage of this approach lies in the fact that the couplings between the system and the detector are sequential, and the measure on the detector is performed only at the end of the procedure.
Therefore, even if the quantum circuit must be iterated to obtain the desired statistical accuracy, we extract the derivative information by performing just a single measurement on the detector qubit instead of the two needed with the  \DM approach.
The details of the \QNDM algorithm can be found in \cite{Solinas_2023}. Here, we summarize only the main steps.

Suppose we are interested in the quantum observable  $\Ham$ in Eq. (\ref{eq:M}).
We add an ancillary qubit and implement the system-detector coupling through the operator $U_\pm = \exp\{\pm i \lambda Z_a \otimes \Ham \}$ where $Z_a$ is a Pauli operator acting on the ancillary detector and $\lambda$ is the system-detector coupling constant. We notice that if $\Ham$ is a complex operator, the implementation of operator $U_\pm$ is, in general, very resource-consuming. This might seem a critical drawback, but the \QNDM approach offers a clear and elegant way to bypass this problem, instead of implementing the operator $U_\pm = \exp\{\pm i \lambda Z_a \otimes \sum_{i=1}^J h_i \hat{P}_i\}$, we can implement the product of single Pauli string operators $\prod_i^J\exp\{\pm i \lambda Z_a \otimes h_i\hat{P_i} \}$ without relying on the Trotterization approximation \cite{Hatano_2005}, as demonstrated in \cite{Solinas_2023}.
Considering the described quantum circuit, this corresponds to the following transformation (in the full system and detector Hilbert space)
\begin{equation}
	U_{tot} =  e^{ i \lambda Z_a \otimes \Ham } U(\vec \theta +s \hat{e}_j ) U^\dagger(\vec \theta -s \hat{e}_j ) e^{-i \lambda Z_a \otimes \Ham } U(\vec \theta -s \hat{e}_j )\,.
	\label{eq:U_tot}
\end{equation} 

The initial state for the system plus the detector $\ket{\psi_0}(\ket{0}_D+ \ket{1}_D)/\sqrt{2}$, where a Hadamard gate is applied to the initial state of the detector.
After implementing the transformation (\ref{eq:U_tot}), we measure the accumulated phase on the detector degrees of freedom.
It can be shown \cite{Solinas_2023, solinas2015fulldistribution, solinas2016probing, solinas2021, solinas2022} that this is a quasi-characteristic function $\GFunc$
\begin{equation}
	\GFunc =  \frac{_D{\matrixel{0}{\rho_D^f }{1}}{_D}}
	{
	_D{\matrixel{0}{\rho_D^0 }{1}}{_D}
    }\,,
	\label{eq:G_lambda_def}
\end{equation}
where $\ket{0}_D$ and $\ket{1}_D$ are eigenstates of the detector operator $Z_a$, while $\rho_D^0$ and $\rho_D^f$ are the density matrices of the detector before and after the application of $U_{tot}$, respectively.

Taking the derivatives of $\GFunc$ with respect to $\lambda$ and evaluating them in $\lambda = 0$, we have access to the information of the different moments of the distribution describing the variation of the cost function \cite{Solinas_2023, solinas2015fulldistribution, solinas2016probing, solinas2021, solinas2022}.
Here, we are interested only in the first moment, which directly gives us $g_j$.
Following Refs. \cite{Solinas_2023,minuto2024novel}, it can be shown that the first derivative of $\GFunc$ reads
\begin{eqnarray}
	-i \partial_\lambda \GFunc \Big |_{\lambda=0} 
    &=&  2 \sum_i h_i \Trace_S [U^\dagger(\vec \theta +s \hat{e}_j ) \hat{P_i} U(\vec \theta +s \hat{e}_j ) \rho_S^0  \nonumber  \\
    &&- U^\dagger(\vec \theta-s \hat{e}_j ) \hat{P_i} U(\vec \theta -s \hat{e}_j )\rho_S^0]
    \,,
	\label{eq:first_derivative}
\end{eqnarray}
where $\Trace_S$ denotes the trace over the {\it system} degrees of freedom.
By direct comparison, Eq. (\ref{eq:first_derivative}) is proportional to Eq. (\ref{eq:DM}),
\begin{equation}
    -i \partial_\lambda \GFunc \Big |_{\lambda=0} = 2\sin(s)g_j.
\end{equation}
Thus, both the \QNDM and the  \DM approach can be used to compute the derivative of $f(\vec \theta)$. 

As discussed in previous works \cite{solinas2021, solinas2022, Solinas_2023,minuto2024novel}, it is possible to approximate the derivative of the quasi-characteristic function at the linear order by opportunity choosing the value of the coupling $\lambda$.
As an estimate, this approximation holds if $\lambda \sum_{i=1}^J h_i <1$ (see Ref. \cite{minuto2024novel} for details).

The phase accumulated by the detector can be measured with interferometric techniques \cite{solinas2021, solinas2022, Solinas_2023, minuto2024novel}.

\section{Numerical simulations}
\label{sec:simulations}

To evaluate and compare the performance of the  \DM and \QNDM approaches, we examined two distinct classes of problems. 
In both cases, the objective is to employ a variational algorithm to determine the minimum energy of the system. 

The first involves physically inspired cases: we focus on the \H2~and \LiH~molecules. 
Despite the small size of these molecules, the computational is resource consuming so the complexity is reduced by approximating the Hamiltonian using underlying symmetries, focusing on low-energy states and considering only the interactions of electrons in the outermost orbitals. Based on these physical approximations, the problem is mapped onto a quantum computer using two-level quantum states (qubits) \cite{qiskit2024, Hamlib}. In general, the mapping process is complex, but we used a Hamiltonian database called \texttt{Hamlib} \cite{Hamlib}, which contains all the necessary information about the physical molecule we wish to simulate. By specifying the number of qubits required, we can directly access the corresponding approximated Hamiltonian in a ready-to-use form for quantum computers, i.e. providing an Hamiltonian written as a sum of Pauli string, such as in Eq. \eqref{eq:M}.

The second problem we study is a more abstract. While keeping the Hamiltonian structure of Eq. \eqref{eq:M}, we increase the number of Pauli strings $J$.
The coefficients $h_i$ are drawn from a Gaussian distribution. 
The freedom in the choice of $J$ allows us to explore different simulation regimes and determine the scaling of computational resources required as a function of the system's complexity.

For all simulations, we use the \texttt{Qiskit} framework with its noiseless simulator \cite{qiskit2024}. This allows us to focus on the ideal performance of the algorithms without the complications of noise from real quantum hardware, and providing a clear benchmark for evaluating the capabilities of the two approaches. 

The simulations are performed as follows. We select a random initial point in the parameter space and run the gradient descent algorithm for a maximum of $10^3$ iterations.
We have verified that this is sufficient to ensure the algorithm's convergence in the tested cases.
We repeat the simulations for $10$ different initial points to avoid any bias introduced by the choice of initial parameters $\vec{\theta}$.

This procedure also allows us to numerically estimate the uncertainty in the minimization process as follow.
For every optimization iteration, we have calculated the average and the standard deviation of the energy data obtained from different starting point in the parameter space. 
The first determinate the convergence curves while the second determine the curve uncertainty regions.


\subsection{\H2 molecule}
\label{sec:H2}

\begin{figure}
    \begin{center}
        \includegraphics[width=1.1\linewidth]{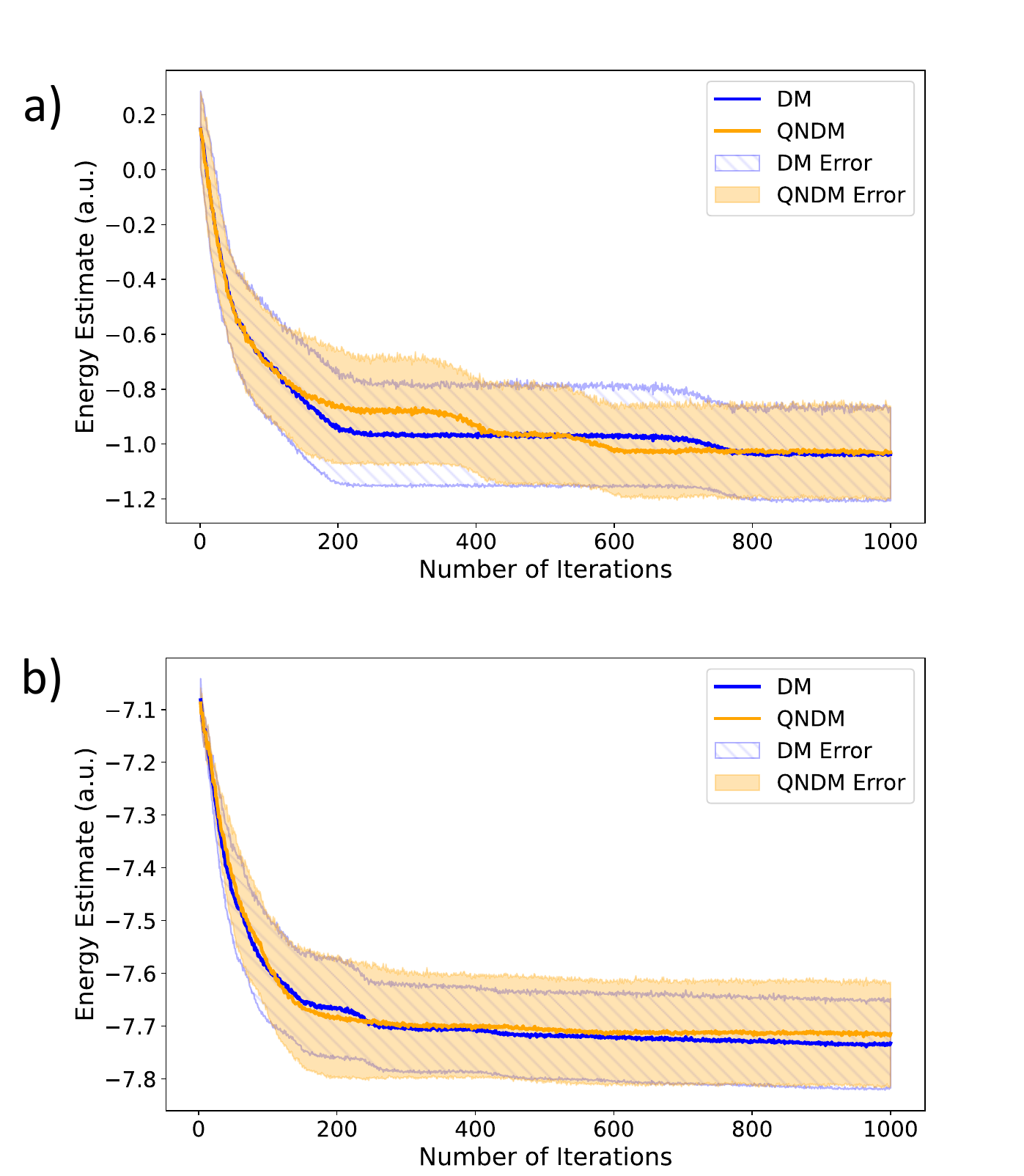}
    \end{center}
    \caption{Energy minimization results for the \H2 (panel a)) and \LiH (panel b)) molecules, using the  \DM (blue) and \QNDM  approaches (orange). 
    For both pictures, the solid lines represent the average convergence curve, i.e., averaged over all the simulation starting from different initial points in the parameter space. 
    The shaded regions represent the statistical uncertainty obtained from different initial points in the parameter space.
    For both simulations, we used $N_{\text{shots}} = 10^3$ and $L = 5$. For the simulation of \H2 in a), we used the gradient descent optimizer [see Eq. \eqref{eq:gradient_descent}] with a learning rate of $\eta = 0.1$ and $\lambda = 0.1$, while for  the simulation of \LiH in b), the same optimizer was used with a learning rate of $\eta = 0.5$ and $\lambda = 0.4$.
    }
    \label{fig:H2_and_LiH_results}
\end{figure}

As a first example, we examined the simplest possible molecule, namely the \H2 molecule. The energy minimization results obtained using both the \DM and \QNDM approaches are shown in Fig.~\ref{fig:H2_and_LiH_results} a). The Hamiltonian contained in the \texttt{Hamlib}~\cite{Hamlib} database, consists of $J = 15$ terms and requires $n = 4$ qubits.

For these simulations, we used $N_{\text{shots}} = 10^3$, $L = 5$, and a gradient descent optimizer [see Eq.~\eqref{eq:gradient_descent}] with a learning rate of $\eta = 0.1$. In the \QNDM approach, we set $\lambda = 0.1$. 
The solid lines denote the average convergence curve, while the shaded regions indicate the statistical uncertainty derived from different initial parameter space points.
The uncertainty region can be reduced by increasing the number of shots (which reduces the statistical uncertainty of the single simulation) or increasing the depth of the quantum circuits by increasing the number of layers (which allows a more precise optimization).

As can be seen in Fig.~\ref{fig:H2_and_LiH_results} a), the two approaches converge towards the minimum energy in a similar way and with a similar uncertainty, i.e. they converge with roughly the same number of iterations and with the same statistical uncertainty.
The convergence occurs within $10^3$ iterations.

Even if the convergence curve is the same for the two approaches, to quantity the simulation resources, we must count for the the different number logical quantum gates needed to calculate the gradient of the energy.
We have that for the  \DM approach the total number of logical quantum gates is $R_{DM} = 2.14\cdot 10^{7}$, while for the \QNDM  one we have $R_{QNDM} =6.2\cdot 10^6$. This represents a practical reduction of approximately $(R_{DM}-R_{QNDM})/R_{DM} = 70\%$ in the number of gates required, highlighting the significant efficiency gain achieved by the \QNDM  method.

The limited dimension of the molecule allows us to perform different tests.
With the above simulating parameters we are in the regimes in which $k \ll nJ$, since $k = 35$, $n=4$ and $J = 15$ \cite{Solinas_2023}.
Especially for larger systems, this regimes might be of limited interest.
In fact, with the increase of the system complexity, we must increase the number of qubits $n$ used in the simulations.
The dimension of the parameter space and the number of layers needed to run the simulations increases exponentially with $n$ \cite{McClean2018, Solinas_2023}.
This trend is unavoidable if we want to maintain the expressivity of the quantum circuit and accurately determine the ground state energy.
As a consequence, the number of logical operation $k$ must increase and, for large system, we are likely to enter the $k \gg n J$.
For example, it has been estimated \cite{Wecker2015} that for the simulations of medium complex molecules   $k\approx 10^9-10^{10}$, $n \approx 10^2-10^3$ (that is the dimension of quantum computer available in the next years) and $J > 10^3$. With these numbers, $k \gg n J$.

Due to its simplicity, this regime is accessible in the \H2 molecule and, to reach it, we have increased the number of layers to $L = 60$, so that $k = 420$ and $nJ \sim 40$. In this regime, the advantage of \QNDM approaches compared to the  \DM one get more relevant (as predicted in Ref. \cite{Solinas_2023}) since for the  \DM method we have a cost of roughly $\sim 3.03\cdot 10^{12}$, while for the \QNDM  approach the total computatinal cost stands at $\sim 2.99 \cdot 10^{11}$, which represents a practical reduction of approximately $70\%$ in the number of gates required for the whole optimization process.


\subsection{\LiH molecule}
\label{sec:Li2}

As a second example, we examine the \LiH molecule. The complexity of this system allows for varying degrees of approximation within the \texttt{Hamlib} database \cite{Hamlib}. By selecting the number of qubits $n$ for the simulation, different approximate representations of the molecular Hamiltonian can be obtained. As $n$ increases, both the number of Pauli terms $J$ in the Hamiltonian and the accuracy of the approximation improve, albeit at the cost of increased computational resources. In order to balance the trade-off between accuracy and computational cost, we selected $n=10$ qubits for the simulation. This choice corresponds to a Hamiltonian with $J \approx 300$ Pauli strings. The results of the energy minimization are shown in Fig. \ref{fig:H2_and_LiH_results} b). The number of shots for a single point is $N_{\text{shots}} = 10^3$. The gradient descent optimizer was used with a learning rate of $\eta = 0.5$, and for the \QNDM  approach, we chose $\lambda = 0.4$.
As above, the solid lines represent the average convergence curve, while the shaded regions represent the statistical uncertainty obtained from $N = 10$ different initial points in the parameter space.
 
Also in this case, we can conclude that the \QNDM and  \DM approaches share a very similar behaviour both in terms of convergence and uncertainty. 
However, taking in consideration this, we can estimate the total resources for \QNDM as $\sim 5.46\cdot 10^{11}$ compared to $\sim 2.62\cdot 10^{12}$ needed for \DM, showing a percentage reduction of $\sim79\%$ in terms of the resources needed. This confirms the advantage of \QNDM approach in terms of resources.

In addition to the simulations presented, we have performed the same analysis for the \OH~and \Li~ molecules (not shown) with similar results and conclusions.

\begin{figure}
    \begin{center}
    \includegraphics[width=1.1\linewidth]{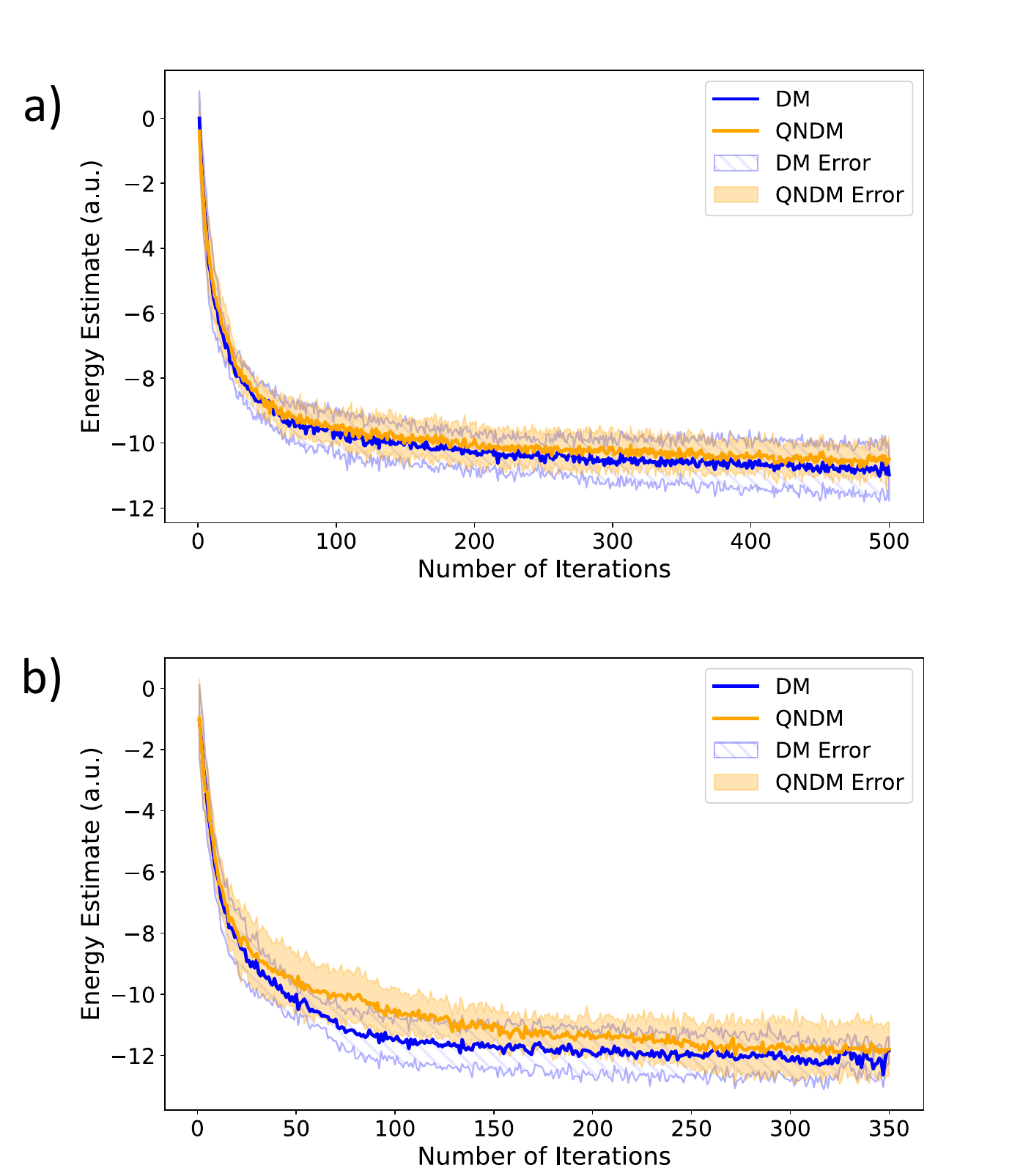}
       \end{center}
    \caption{
    Convergence to the minimum energy for large Hamiltonians with $J = 750$ a) and $J = 1000$ b). The simulations are performed with $n=10$ qubits and with random coefficient $h_i$ drawn from a Gaussian distribution of $\mu = 1$ and $\sigma=0.1$. The gradient descent optimizer is used with a learning rate of $\eta = 0.05$ and, for the \QNDM  approach, $\lambda = 0.01$. The number of shots for a single point is $N_{\text{shots}} = 10^3$ and $L = 5$.}
    \label{fig:large_hamiltonian}
\end{figure}


\subsection{Large Hamiltonian}
\label{sec:Large_H}

In the \texttt{Hamlib}  repository the increase in the Hamiltonian complexity, i.e., increase in $J$, is always associated to an increase of the number of qubits.
Unfortunately, it turns out that an increase of $n$ quickly lead to a saturation of the simulation resources and times, especially, if we want to simulate the full convergence process and have sufficient statistical analysis.
For this reason, to validate the previous results of the \QNDM approach respect to the  \DM one, we focus on a different (but still physical) system. 

We consider a problem where the Hamiltonian consists of a large and variable number of Pauli strings $J$. The Hamiltonian has the same form of Eq. \eqref{eq:M}, but the coefficients of the Pauli strings $h_i$ are generated randomly from a Gaussian distribution with mean $\mu = 1$ and standard deviation $\sigma = 0.1$. The simulations are performed for $n=10$ qubits and the number of shots for a single point is $N_{\text{shots}} = 10^3$ and $L = 5$.
For both cases, we considered the gradient descent optimizer with a learning rate of $\eta = 0.05$ and $\lambda = 0.01$ for the \QNDM  approach.

The results for $J=750$ and $J=1000$ are shown in Fig. \ref{fig:large_hamiltonian}. 
The two approaches are similar in number of iterations to reach the convergence and statistical uncertainty.

\begin{figure}
    \centering
    \includegraphics[width=1.1\linewidth]{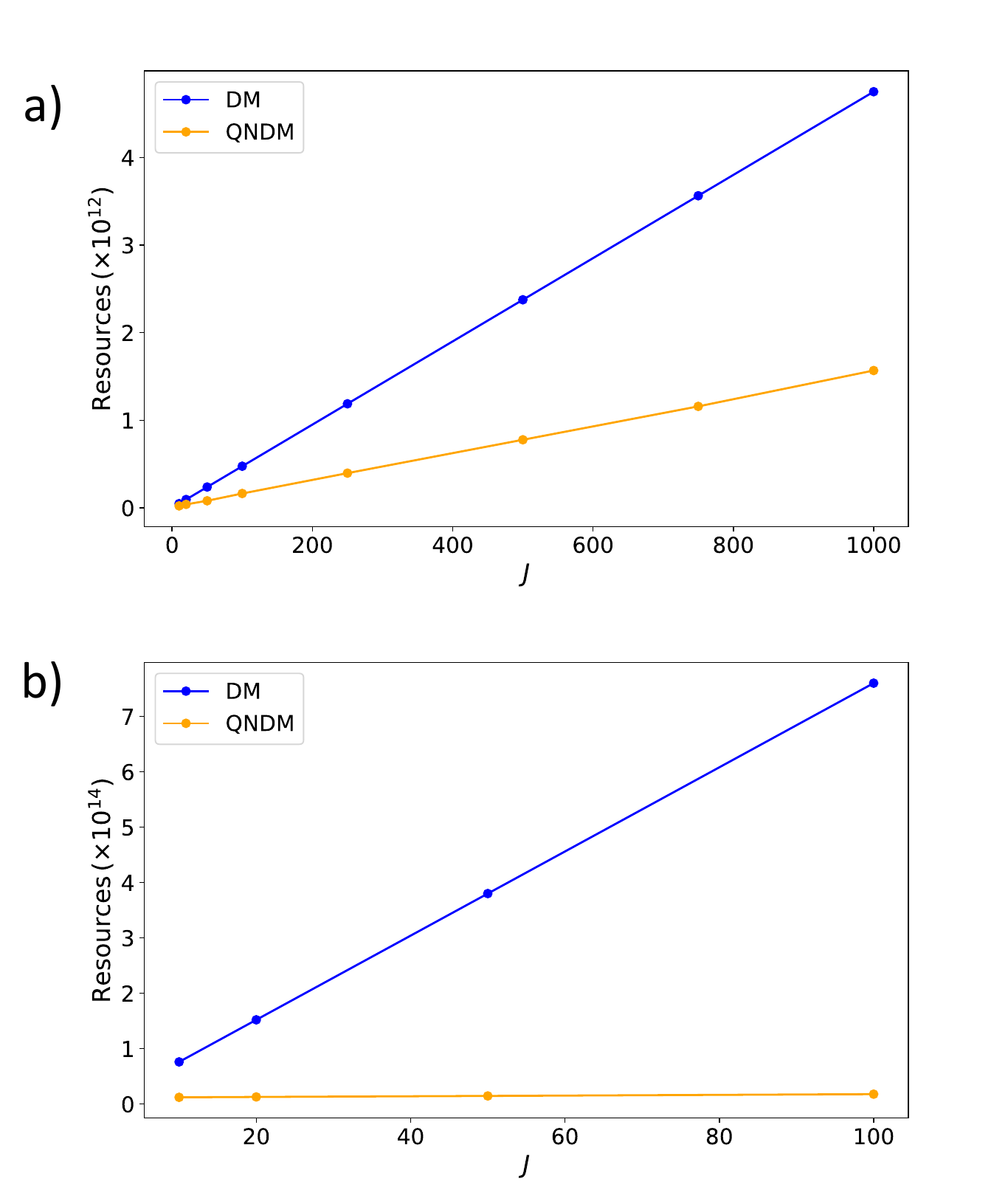}
    \caption{Total resources scaling as a function of $J$
    a) Regime $k \ll nJ$. b) Regime $k \gg nJ$.
    The simulations are performed with $n = 10$ qubits, and for the a) we have $L = 5$, while for the b) we have the same number of qubits, but $L = 200$.
    }
    \label{fig:resource_scaling}
\end{figure}

As above, the relevant parameter is the total number of resources needed of the simulations.
The possibility to change $J$ allows us to study the two relevant regimes $k \ll nJ$ and $k \gg nJ$.
The simulations in Fig. \ref{fig:large_hamiltonian} are in the regime $k \ll nJ$ and the corresponding resource count as a function of $J$ is plotted Fig. \ref{fig:resource_scaling} a).
The \QNDM approach has an advantage in terms of reduced number of logical operation needed so that, for $J=1000$, the \DM approach needs more than four time the resources needed by the \QNDM one.

Because of limitations in the computing times, these simulations have been preformed with a limited number of layers, namely $L = 5$. Even if the full simulations are beyond the reach of our computationally resources, it is interesting to estimate the resources in the opposite regime, i.e., $k \gg nJ$, since, as discussed Sec. \ref{sec:H2}, this is likely to be the more interesting one for complex physical and chemical systems.

The results are shown in Fig. \ref{fig:resource_scaling} b).
Here, we are supposing to simulate a Hamiltonian with limited number of Pauli strings $J$ but want to increase the expressivity of the quantum circuits and obtain more accurate energy minima.
As discussed in Ref. \cite{Solinas_2023}, in this case we have a linear advantage in $J$ of the \QNDM  versus the  \DM approach.
Indeed, while the \QNDM  is almost constant and requires a number of logical operation of the order of roughly $1.7\cdot 10^{11}$, the resources for  \DM increase linearly with $J$ to reach roughly $7 \cdot 10^{14}$.

\section{Conclusions}
\label{sec:conclusions}

We have performed an in-depth study of the performance of the Variational quantum algorithms for minimizing the energy of a quantum system.
We focused on hybrid models where the information extracted with a quantum computer is then elaborated from a classical computer and then fed back to the quantum hardware.
The application of a gradient decent algorithm allows us to reach the minimal energy state within a simulable number of iterations.

Within this framework, we have compared two approaches that differ in the way we extract the information about the gradient of the energy in a point of the parameter space.
The first one, called Direct Measurement, is based on iterated projective measurements of the system \cite{schuld2015, McClean_2016, McClean2018,SchuldPRA2019, Mari2021, Cerezo2021}
The second one, leverages the so-called quantum non-demolition measurement scheme, exploits an additional quantum detector and stores the information about the gradient in the detector phase, which is eventually measured \cite{Solinas_2023, minuto2024novel}.

As physical models, we have used the approximated description of \H2 and \LiH molecules.
Starting from the \texttt{Hamlib} database \cite{Hamlib}, these are are directly simulated in quantum computers.

We have shown that both  \DM and \QNDM  methods reach the minimum energy state within a comparable number of iteration, and also with a similar statistical uncertainty in the energy minimization.

However, since the \QNDM  approach is more efficient in evaluating the gradient, we observe a reduction of the overall resources.
For the chemical molecules studies, this corresponds to a cost reduction of roughly $70\%$ for the \H2 complete optimization process, and roughly $79\%$ for the \LiH one.

We have also consider the case of Hamiltonians composed of a large number of Pauli strings $J$, where the Pauli strings are extracted randomly, and the associated coefficients are drawn from a Gaussian distribution.
In this case, we were able to compare the two approaches in different regimes $k \ll nJ$ and $k \gg nJ$.
In both we observe a smaller number of resources needed for the \QNDM optimization.
In the second regime $k \gg nJ$, the advantage of the \QNDM approach is more evident, since the resources for \DM increase linearly with $J$ to reach, for $J = 1000$, roughly $7 \cdot 10^{14}$, while the \QNDM  is almost constant and requires a number of logical operation of the order of roughly $1.7\cdot 10^{11}$.

Our results suggest that the \QNDM  can be a viable alternative (and, probably, the best way) to implement the variational quantum algorithm for the energy, or in general cost function, minimization tasks.
We expect this advantage to increase with the complexity of the simulated system and, therefore, its interest.


\section*{Acknowledgements}
The authors acknowledge financial support from INFN. This work was carried out while GM was enrolled in the Italian National Doctorate on Artificial Intelligence run by Sapienza University of Rome in collaboration with the Dept. of Informatics, Bioengineering, Robotics, and Systems Engineering, Polytechnic School of Genova.




\begin{thebibliography}{31}
    \expandafter\ifx\csname natexlab\endcsname\relax\def\natexlab#1{#1}\fi
    \expandafter\ifx\csname bibnamefont\endcsname\relax
      \def\bibnamefont#1{#1}\fi
    \expandafter\ifx\csname bibfnamefont\endcsname\relax
      \def\bibfnamefont#1{#1}\fi
    \expandafter\ifx\csname citenamefont\endcsname\relax
      \def\citenamefont#1{#1}\fi
    \expandafter\ifx\csname url\endcsname\relax
      \def\url#1{\texttt{#1}}\fi
    \expandafter\ifx\csname urlprefix\endcsname\relax\def\urlprefix{URL }\fi
    \providecommand{\bibinfo}[2]{#2}
    \providecommand{\eprint}[2][]{\url{#2}}
    
    \bibitem[{\citenamefont{Feynman}(1982)}]{feynman1982}
    \bibinfo{author}{\bibfnamefont{R.~P.} \bibnamefont{Feynman}}, \bibinfo{journal}{International journal of theoretical physics} \textbf{\bibinfo{volume}{21}}, \bibinfo{pages}{467} (\bibinfo{year}{1982}).
    
    \bibitem[{\citenamefont{Peruzzo et~al.}(2014)\citenamefont{Peruzzo, McClean, Shadbolt, Yung, Zhou, Love, Aspuru-Guzik, and O'Brien}}]{Peruzzo_2014}
    \bibinfo{author}{\bibfnamefont{A.}~\bibnamefont{Peruzzo}}, \bibinfo{author}{\bibfnamefont{J.}~\bibnamefont{McClean}}, \bibinfo{author}{\bibfnamefont{P.}~\bibnamefont{Shadbolt}}, \bibinfo{author}{\bibfnamefont{M.-H.} \bibnamefont{Yung}}, \bibinfo{author}{\bibfnamefont{X.-Q.} \bibnamefont{Zhou}}, \bibinfo{author}{\bibfnamefont{P.~J.} \bibnamefont{Love}}, \bibinfo{author}{\bibfnamefont{A.}~\bibnamefont{Aspuru-Guzik}}, \bibnamefont{and} \bibinfo{author}{\bibfnamefont{J.~L.} \bibnamefont{O'Brien}}, \bibinfo{journal}{Nature Communications} \textbf{\bibinfo{volume}{5}} (\bibinfo{year}{2014}), \urlprefix\url{https://doi.org/10.1038%2Fncomms5213}.
    
    \bibitem[{\citenamefont{Maria~Schuld and Petruccione}(2015)}]{schuld2015}
    \bibinfo{author}{\bibfnamefont{I.~S.} \bibnamefont{Maria~Schuld}} \bibnamefont{and} \bibinfo{author}{\bibfnamefont{F.}~\bibnamefont{Petruccione}}, \bibinfo{journal}{Contemporary Physics} \textbf{\bibinfo{volume}{56}}, \bibinfo{pages}{172} (\bibinfo{year}{2015}), \eprint{https://doi.org/10.1080/00107514.2014.964942}, \urlprefix\url{https://doi.org/10.1080/00107514.2014.964942}.
    
    \bibitem[{\citenamefont{McClean et~al.}(2016)\citenamefont{McClean, Romero, Babbush, and Aspuru-Guzik}}]{McClean_2016}
    \bibinfo{author}{\bibfnamefont{J.~R.} \bibnamefont{McClean}}, \bibinfo{author}{\bibfnamefont{J.}~\bibnamefont{Romero}}, \bibinfo{author}{\bibfnamefont{R.}~\bibnamefont{Babbush}}, \bibnamefont{and} \bibinfo{author}{\bibfnamefont{A.}~\bibnamefont{Aspuru-Guzik}}, \bibinfo{journal}{New Journal of Physics} \textbf{\bibinfo{volume}{18}}, \bibinfo{pages}{023023} (\bibinfo{year}{2016}), ISSN \bibinfo{issn}{1367-2630}, \urlprefix\url{http://dx.doi.org/10.1088/1367-2630/18/2/023023}.
    
    \bibitem[{\citenamefont{McClean et~al.}(2018)\citenamefont{McClean, Boixo, Smelyanskiy, Babbush, and Neven}}]{McClean2018}
    \bibinfo{author}{\bibfnamefont{J.~R.} \bibnamefont{McClean}}, \bibinfo{author}{\bibfnamefont{S.}~\bibnamefont{Boixo}}, \bibinfo{author}{\bibfnamefont{V.~N.} \bibnamefont{Smelyanskiy}}, \bibinfo{author}{\bibfnamefont{R.}~\bibnamefont{Babbush}}, \bibnamefont{and} \bibinfo{author}{\bibfnamefont{H.}~\bibnamefont{Neven}}, \bibinfo{journal}{Nature Communications} \textbf{\bibinfo{volume}{9}}, \bibinfo{pages}{4812} (\bibinfo{year}{2018}), \urlprefix\url{https://doi.org/10.1038/s41467-018-07090-4}.
    
    \bibitem[{\citenamefont{Schuld et~al.}(2019)\citenamefont{Schuld, Bergholm, Gogolin, Izaac, and Killoran}}]{SchuldPRA2019}
    \bibinfo{author}{\bibfnamefont{M.}~\bibnamefont{Schuld}}, \bibinfo{author}{\bibfnamefont{V.}~\bibnamefont{Bergholm}}, \bibinfo{author}{\bibfnamefont{C.}~\bibnamefont{Gogolin}}, \bibinfo{author}{\bibfnamefont{J.}~\bibnamefont{Izaac}}, \bibnamefont{and} \bibinfo{author}{\bibfnamefont{N.}~\bibnamefont{Killoran}}, \bibinfo{journal}{Phys. Rev. A} \textbf{\bibinfo{volume}{99}}, \bibinfo{pages}{032331} (\bibinfo{year}{2019}), \urlprefix\url{https://link.aps.org/doi/10.1103/PhysRevA.99.032331}.
    
    \bibitem[{\citenamefont{Mari et~al.}(2021)\citenamefont{Mari, Bromley, and Killoran}}]{Mari2021}
    \bibinfo{author}{\bibfnamefont{A.}~\bibnamefont{Mari}}, \bibinfo{author}{\bibfnamefont{T.~R.} \bibnamefont{Bromley}}, \bibnamefont{and} \bibinfo{author}{\bibfnamefont{N.}~\bibnamefont{Killoran}}, \bibinfo{journal}{Phys. Rev. A} \textbf{\bibinfo{volume}{103}}, \bibinfo{pages}{012405} (\bibinfo{year}{2021}), \urlprefix\url{https://link.aps.org/doi/10.1103/PhysRevA.103.012405}.
    
    \bibitem[{\citenamefont{Cerezo et~al.}(2021)\citenamefont{Cerezo, Arrasmith, Babbush, Benjamin, Endo, Fujii, McClean, Mitarai, Yuan, Cincio et~al.}}]{Cerezo2021}
    \bibinfo{author}{\bibfnamefont{M.}~\bibnamefont{Cerezo}}, \bibinfo{author}{\bibfnamefont{A.}~\bibnamefont{Arrasmith}}, \bibinfo{author}{\bibfnamefont{R.}~\bibnamefont{Babbush}}, \bibinfo{author}{\bibfnamefont{S.~C.} \bibnamefont{Benjamin}}, \bibinfo{author}{\bibfnamefont{S.}~\bibnamefont{Endo}}, \bibinfo{author}{\bibfnamefont{K.}~\bibnamefont{Fujii}}, \bibinfo{author}{\bibfnamefont{J.~R.} \bibnamefont{McClean}}, \bibinfo{author}{\bibfnamefont{K.}~\bibnamefont{Mitarai}}, \bibinfo{author}{\bibfnamefont{X.}~\bibnamefont{Yuan}}, \bibinfo{author}{\bibfnamefont{L.}~\bibnamefont{Cincio}}, \bibnamefont{et~al.}, \bibinfo{journal}{Nature Reviews Physics} \textbf{\bibinfo{volume}{3}}, \bibinfo{pages}{625} (\bibinfo{year}{2021}), \urlprefix\url{https://doi.org/10.1038/s42254-021-00348-9}.
    
    \bibitem[{\citenamefont{Solinas et~al.}(2023)\citenamefont{Solinas, Caletti, and Minuto}}]{Solinas_2023}
    \bibinfo{author}{\bibfnamefont{P.}~\bibnamefont{Solinas}}, \bibinfo{author}{\bibfnamefont{S.}~\bibnamefont{Caletti}}, \bibnamefont{and} \bibinfo{author}{\bibfnamefont{G.}~\bibnamefont{Minuto}}, \bibinfo{journal}{The European Physical Journal D} \textbf{\bibinfo{volume}{77}} (\bibinfo{year}{2023}), ISSN \bibinfo{issn}{1434-6079}, \urlprefix\url{http://dx.doi.org/10.1140/epjd/s10053-023-00648-y}.
    
    \bibitem[{\citenamefont{Minuto et~al.}(2024)\citenamefont{Minuto, Caletti, and Solinas}}]{minuto2024novel}
    \bibinfo{author}{\bibfnamefont{G.}~\bibnamefont{Minuto}}, \bibinfo{author}{\bibfnamefont{S.}~\bibnamefont{Caletti}}, \bibnamefont{and} \bibinfo{author}{\bibfnamefont{P.}~\bibnamefont{Solinas}}, \bibinfo{journal}{arXiv preprint arXiv:2404.02245}  (\bibinfo{year}{2024}).
    
    \bibitem[{\citenamefont{Braginsky et~al.}(1980)\citenamefont{Braginsky, Vorontsov, and Thorne}}]{Braginsky1980}
    \bibinfo{author}{\bibfnamefont{V.~B.} \bibnamefont{Braginsky}}, \bibinfo{author}{\bibfnamefont{Y.~I.} \bibnamefont{Vorontsov}}, \bibnamefont{and} \bibinfo{author}{\bibfnamefont{K.~S.} \bibnamefont{Thorne}}, \bibinfo{journal}{Science} \textbf{\bibinfo{volume}{209}}, \bibinfo{pages}{547} (\bibinfo{year}{1980}), \eprint{https://www.science.org/doi/pdf/10.1126/science.209.4456.547}, \urlprefix\url{https://www.science.org/doi/abs/10.1126/science.209.4456.547}.
    
    \bibitem[{\citenamefont{Braginski{\u \i} et~al.}(1992)\citenamefont{Braginski{\u \i}, Khalili, and Thorne}}]{BraginskyBook}
    \bibinfo{author}{\bibfnamefont{V.~B.} \bibnamefont{Braginski{\u \i}}}, \bibinfo{author}{\bibfnamefont{F.~Y.} \bibnamefont{Khalili}}, \bibnamefont{and} \bibinfo{author}{\bibfnamefont{K.~S.} \bibnamefont{Thorne}}, \emph{\bibinfo{title}{Quantum measurement}} (\bibinfo{publisher}{Cambridge University Press}, \bibinfo{address}{Cambridge [England] ;}, \bibinfo{year}{1992}), ISBN \bibinfo{isbn}{052141928X}.
    
    \bibitem[{\citenamefont{Caves}(1980)}]{Caves1980}
    \bibinfo{author}{\bibfnamefont{C.~M.} \bibnamefont{Caves}}, \bibinfo{journal}{Phys. Rev. Lett.} \textbf{\bibinfo{volume}{45}}, \bibinfo{pages}{75} (\bibinfo{year}{1980}), \urlprefix\url{https://link.aps.org/doi/10.1103/PhysRevLett.45.75}.
    
    \bibitem[{\citenamefont{Solinas and Gasparinetti}(2015)}]{solinas2015fulldistribution}
    \bibinfo{author}{\bibfnamefont{P.}~\bibnamefont{Solinas}} \bibnamefont{and} \bibinfo{author}{\bibfnamefont{S.}~\bibnamefont{Gasparinetti}}, \bibinfo{journal}{Phys. Rev. E} \textbf{\bibinfo{volume}{92}}, \bibinfo{pages}{042150} (\bibinfo{year}{2015}), \urlprefix\url{http://link.aps.org/doi/10.1103/PhysRevE.92.042150}.
    
    \bibitem[{\citenamefont{Solinas and Gasparinetti}(2016)}]{solinas2016probing}
    \bibinfo{author}{\bibfnamefont{P.}~\bibnamefont{Solinas}} \bibnamefont{and} \bibinfo{author}{\bibfnamefont{S.}~\bibnamefont{Gasparinetti}}, \bibinfo{journal}{Phys. Rev. A} \textbf{\bibinfo{volume}{94}}, \bibinfo{pages}{052103} (\bibinfo{year}{2016}), \urlprefix\url{https://link.aps.org/doi/10.1103/PhysRevA.94.052103}.
    
    \bibitem[{\citenamefont{Solinas et~al.}(2021)\citenamefont{Solinas, Amico, and Zangh\`{\i}}}]{solinas2021}
    \bibinfo{author}{\bibfnamefont{P.}~\bibnamefont{Solinas}}, \bibinfo{author}{\bibfnamefont{M.}~\bibnamefont{Amico}}, \bibnamefont{and} \bibinfo{author}{\bibfnamefont{N.}~\bibnamefont{Zangh\`{\i}}}, \bibinfo{journal}{Phys. Rev. A} \textbf{\bibinfo{volume}{103}}, \bibinfo{pages}{L060202} (\bibinfo{year}{2021}), \urlprefix\url{https://link.aps.org/doi/10.1103/PhysRevA.103.L060202}.
    
    \bibitem[{\citenamefont{Solinas et~al.}(2022)\citenamefont{Solinas, Amico, and Zangh\`{\i}}}]{solinas2022}
    \bibinfo{author}{\bibfnamefont{P.}~\bibnamefont{Solinas}}, \bibinfo{author}{\bibfnamefont{M.}~\bibnamefont{Amico}}, \bibnamefont{and} \bibinfo{author}{\bibfnamefont{N.}~\bibnamefont{Zangh\`{\i}}}, \bibinfo{journal}{Phys. Rev. A} \textbf{\bibinfo{volume}{105}}, \bibinfo{pages}{032606} (\bibinfo{year}{2022}), \urlprefix\url{https://link.aps.org/doi/10.1103/PhysRevA.105.032606}.
    
    \bibitem[{\citenamefont{McArdle et~al.}(2020)\citenamefont{McArdle, Endo, Aspuru-Guzik, Benjamin, and Yuan}}]{McArdle2020}
    \bibinfo{author}{\bibfnamefont{S.}~\bibnamefont{McArdle}}, \bibinfo{author}{\bibfnamefont{S.}~\bibnamefont{Endo}}, \bibinfo{author}{\bibfnamefont{A.}~\bibnamefont{Aspuru-Guzik}}, \bibinfo{author}{\bibfnamefont{S.~C.} \bibnamefont{Benjamin}}, \bibnamefont{and} \bibinfo{author}{\bibfnamefont{X.}~\bibnamefont{Yuan}}, \bibinfo{journal}{Rev. Mod. Phys.} \textbf{\bibinfo{volume}{92}}, \bibinfo{pages}{015003} (\bibinfo{year}{2020}), \urlprefix\url{https://link.aps.org/doi/10.1103/RevModPhys.92.015003}.
    
    \bibitem[{\citenamefont{Tilly et~al.}(2022)\citenamefont{Tilly, Chen, Cao, Picozzi, Setia, Li, Grant, Wossnig, Rungger, Booth et~al.}}]{TILLY20221}
    \bibinfo{author}{\bibfnamefont{J.}~\bibnamefont{Tilly}}, \bibinfo{author}{\bibfnamefont{H.}~\bibnamefont{Chen}}, \bibinfo{author}{\bibfnamefont{S.}~\bibnamefont{Cao}}, \bibinfo{author}{\bibfnamefont{D.}~\bibnamefont{Picozzi}}, \bibinfo{author}{\bibfnamefont{K.}~\bibnamefont{Setia}}, \bibinfo{author}{\bibfnamefont{Y.}~\bibnamefont{Li}}, \bibinfo{author}{\bibfnamefont{E.}~\bibnamefont{Grant}}, \bibinfo{author}{\bibfnamefont{L.}~\bibnamefont{Wossnig}}, \bibinfo{author}{\bibfnamefont{I.}~\bibnamefont{Rungger}}, \bibinfo{author}{\bibfnamefont{G.~H.} \bibnamefont{Booth}}, \bibnamefont{et~al.}, \bibinfo{journal}{Physics Reports} \textbf{\bibinfo{volume}{986}}, \bibinfo{pages}{1} (\bibinfo{year}{2022}), ISSN \bibinfo{issn}{0370-1573}, \bibinfo{note}{the Variational Quantum Eigensolver: a review of methods and best practices}, \urlprefix\url{https://www.sciencedirect.com/science/article/pii/S0370157322003118}.
    
    \bibitem[{\citenamefont{Nielsen and Chuang}(2010)}]{nielsen-chuang_book}
    \bibinfo{author}{\bibfnamefont{M.~A.} \bibnamefont{Nielsen}} \bibnamefont{and} \bibinfo{author}{\bibfnamefont{I.~L.} \bibnamefont{Chuang}}, \emph{\bibinfo{title}{Quantum computation and quantum information}} (\bibinfo{publisher}{Cambridge University Press, Cambridge, UK}, \bibinfo{year}{2010}).
    
    \bibitem[{\citenamefont{Kingma and Ba}(2017)}]{kingma2017adam}
    \bibinfo{author}{\bibfnamefont{D.~P.} \bibnamefont{Kingma}} \bibnamefont{and} \bibinfo{author}{\bibfnamefont{J.}~\bibnamefont{Ba}}, \emph{\bibinfo{title}{Adam: A method for stochastic optimization}} (\bibinfo{year}{2017}), \eprint{1412.6980}.
    
    \bibitem[{\citenamefont{K{\"{u}}bler et~al.}(2020)\citenamefont{K{\"{u}}bler, Arrasmith, Cincio, and Coles}}]{Kubler2020adaptiveoptimizer}
    \bibinfo{author}{\bibfnamefont{J.~M.} \bibnamefont{K{\"{u}}bler}}, \bibinfo{author}{\bibfnamefont{A.}~\bibnamefont{Arrasmith}}, \bibinfo{author}{\bibfnamefont{L.}~\bibnamefont{Cincio}}, \bibnamefont{and} \bibinfo{author}{\bibfnamefont{P.~J.} \bibnamefont{Coles}}, \bibinfo{journal}{{Quantum}} \textbf{\bibinfo{volume}{4}}, \bibinfo{pages}{263} (\bibinfo{year}{2020}), ISSN \bibinfo{issn}{2521-327X}, \urlprefix\url{https://doi.org/10.22331/q-2020-05-11-263}.
    
    \bibitem[{\citenamefont{Sweke et~al.}(2020)\citenamefont{Sweke, Wilde, Meyer, Schuld, Faehrmann, Meynard-Piganeau, and Eisert}}]{Sweke2020stochasticgradient}
    \bibinfo{author}{\bibfnamefont{R.}~\bibnamefont{Sweke}}, \bibinfo{author}{\bibfnamefont{F.}~\bibnamefont{Wilde}}, \bibinfo{author}{\bibfnamefont{J.}~\bibnamefont{Meyer}}, \bibinfo{author}{\bibfnamefont{M.}~\bibnamefont{Schuld}}, \bibinfo{author}{\bibfnamefont{P.~K.} \bibnamefont{Faehrmann}}, \bibinfo{author}{\bibfnamefont{B.}~\bibnamefont{Meynard-Piganeau}}, \bibnamefont{and} \bibinfo{author}{\bibfnamefont{J.}~\bibnamefont{Eisert}}, \bibinfo{journal}{{Quantum}} \textbf{\bibinfo{volume}{4}}, \bibinfo{pages}{314} (\bibinfo{year}{2020}), ISSN \bibinfo{issn}{2521-327X}, \urlprefix\url{https://doi.org/10.22331/q-2020-08-31-314}.
    
    \bibitem[{\citenamefont{Li et~al.}(2017)\citenamefont{Li, Yang, Peng, and Sun}}]{Li2017}
    \bibinfo{author}{\bibfnamefont{J.}~\bibnamefont{Li}}, \bibinfo{author}{\bibfnamefont{X.}~\bibnamefont{Yang}}, \bibinfo{author}{\bibfnamefont{X.}~\bibnamefont{Peng}}, \bibnamefont{and} \bibinfo{author}{\bibfnamefont{C.-P.} \bibnamefont{Sun}}, \bibinfo{journal}{Phys. Rev. Lett.} \textbf{\bibinfo{volume}{118}}, \bibinfo{pages}{150503} (\bibinfo{year}{2017}), \urlprefix\url{https://link.aps.org/doi/10.1103/PhysRevLett.118.150503}.
    
    \bibitem[{\citenamefont{Mitarai et~al.}(2018)\citenamefont{Mitarai, Negoro, Kitagawa, and Fujii}}]{Mitarai2018}
    \bibinfo{author}{\bibfnamefont{K.}~\bibnamefont{Mitarai}}, \bibinfo{author}{\bibfnamefont{M.}~\bibnamefont{Negoro}}, \bibinfo{author}{\bibfnamefont{M.}~\bibnamefont{Kitagawa}}, \bibnamefont{and} \bibinfo{author}{\bibfnamefont{K.}~\bibnamefont{Fujii}}, \bibinfo{journal}{Phys. Rev. A} \textbf{\bibinfo{volume}{98}}, \bibinfo{pages}{032309} (\bibinfo{year}{2018}), \urlprefix\url{https://link.aps.org/doi/10.1103/PhysRevA.98.032309}.
    
    \bibitem[{\citenamefont{Mitarai and Fujii}(2019)}]{Mitarai2019}
    \bibinfo{author}{\bibfnamefont{K.}~\bibnamefont{Mitarai}} \bibnamefont{and} \bibinfo{author}{\bibfnamefont{K.}~\bibnamefont{Fujii}}, \bibinfo{journal}{Phys. Rev. Research} \textbf{\bibinfo{volume}{1}}, \bibinfo{pages}{013006} (\bibinfo{year}{2019}), \urlprefix\url{https://link.aps.org/doi/10.1103/PhysRevResearch.1.013006}.
    
    \bibitem[{\citenamefont{Ruder}(2017)}]{ruder2017overviewgradientdescentoptimization}
    \bibinfo{author}{\bibfnamefont{S.}~\bibnamefont{Ruder}}, \emph{\bibinfo{title}{An overview of gradient descent optimization algorithms}} (\bibinfo{year}{2017}), \eprint{1609.04747}, \urlprefix\url{https://arxiv.org/abs/1609.04747}.
    
    \bibitem[{\citenamefont{Hatano and Suzuki}(2005)}]{Hatano_2005}
    \bibinfo{author}{\bibfnamefont{N.}~\bibnamefont{Hatano}} \bibnamefont{and} \bibinfo{author}{\bibfnamefont{M.}~\bibnamefont{Suzuki}}, \emph{\bibinfo{title}{Finding Exponential Product Formulas of Higher Orders}} (\bibinfo{publisher}{Springer Berlin Heidelberg}, \bibinfo{year}{2005}), pp. \bibinfo{pages}{37--68}, ISBN \bibinfo{isbn}{9783540315155}, \urlprefix\url{http://dx.doi.org/10.1007/11526216_2}.
    
    \bibitem[{\citenamefont{Javadi-Abhari et~al.}(2024)\citenamefont{Javadi-Abhari, Treinish, Krsulich, Wood, Lishman, Gacon, Martiel, Nation, Bishop, Cross et~al.}}]{qiskit2024}
    \bibinfo{author}{\bibfnamefont{A.}~\bibnamefont{Javadi-Abhari}}, \bibinfo{author}{\bibfnamefont{M.}~\bibnamefont{Treinish}}, \bibinfo{author}{\bibfnamefont{K.}~\bibnamefont{Krsulich}}, \bibinfo{author}{\bibfnamefont{C.~J.} \bibnamefont{Wood}}, \bibinfo{author}{\bibfnamefont{J.}~\bibnamefont{Lishman}}, \bibinfo{author}{\bibfnamefont{J.}~\bibnamefont{Gacon}}, \bibinfo{author}{\bibfnamefont{S.}~\bibnamefont{Martiel}}, \bibinfo{author}{\bibfnamefont{P.~D.} \bibnamefont{Nation}}, \bibinfo{author}{\bibfnamefont{L.~S.} \bibnamefont{Bishop}}, \bibinfo{author}{\bibfnamefont{A.~W.} \bibnamefont{Cross}}, \bibnamefont{et~al.}, \emph{\bibinfo{title}{Quantum computing with {Q}iskit}} (\bibinfo{year}{2024}), \eprint{2405.08810}.
    
    \bibitem[{\citenamefont{Sawaya et~al.}(2024)\citenamefont{Sawaya, Marti-Dafcik, Ho, Tabor, Neira, Magann, Premaratne, Dubey, Matsuura, Bishop et~al.}}]{Hamlib}
    \bibinfo{author}{\bibfnamefont{N.~P.} \bibnamefont{Sawaya}}, \bibinfo{author}{\bibfnamefont{D.}~\bibnamefont{Marti-Dafcik}}, \bibinfo{author}{\bibfnamefont{Y.}~\bibnamefont{Ho}}, \bibinfo{author}{\bibfnamefont{D.~P.} \bibnamefont{Tabor}}, \bibinfo{author}{\bibfnamefont{D.~E.~B.} \bibnamefont{Neira}}, \bibinfo{author}{\bibfnamefont{A.~B.} \bibnamefont{Magann}}, \bibinfo{author}{\bibfnamefont{S.}~\bibnamefont{Premaratne}}, \bibinfo{author}{\bibfnamefont{P.}~\bibnamefont{Dubey}}, \bibinfo{author}{\bibfnamefont{A.}~\bibnamefont{Matsuura}}, \bibinfo{author}{\bibfnamefont{N.}~\bibnamefont{Bishop}}, \bibnamefont{et~al.}, \bibinfo{journal}{Quantum} \textbf{\bibinfo{volume}{8}}, \bibinfo{pages}{1559} (\bibinfo{year}{2024}), ISSN \bibinfo{issn}{2521-327X}, \urlprefix\url{http://dx.doi.org/10.22331/q-2024-12-11-1559}.
    
    \bibitem[{\citenamefont{Wecker et~al.}(2015)\citenamefont{Wecker, Hastings, and Troyer}}]{Wecker2015}
    \bibinfo{author}{\bibfnamefont{D.}~\bibnamefont{Wecker}}, \bibinfo{author}{\bibfnamefont{M.~B.} \bibnamefont{Hastings}}, \bibnamefont{and} \bibinfo{author}{\bibfnamefont{M.}~\bibnamefont{Troyer}}, \bibinfo{journal}{Phys. Rev. A} \textbf{\bibinfo{volume}{92}}, \bibinfo{pages}{042303} (\bibinfo{year}{2015}), \urlprefix\url{https://link.aps.org/doi/10.1103/PhysRevA.92.042303}.
    
\end{thebibliography}
\end{document}